\documentclass[sigconf]{acmart}
\copyrightyear{2023} 
\acmYear{2023} 
\setcopyright{rightsretained} 
\acmConference[ASIA CCS '23]{ACM ASIA Conference on Computer and Communications Security}{July 10--14, 2023}{Melbourne, VIC, Australia}
\acmBooktitle{ACM ASIA Conference on Computer and Communications Security (ASIA CCS '23), July 10--14, 2023, Melbourne, VIC, Australia}
\acmDOI{10.1145/3579856.3582824}
\acmISBN{979-8-4007-0098-9/23/07}
\usepackage{amsmath,amsfonts}
\usepackage{graphicx}
\DeclareMathOperator*{\argmax}{arg\,max}

\usepackage{graphicx}
\usepackage{hyperref}
\usepackage{url}
\usepackage{float}
\usepackage{setspace}
\usepackage{harpoon}
\usepackage{multicol}
\usepackage{multirow}
\usepackage{caption}
\usepackage{hhline}
\usepackage[labelformat=simple]{subcaption}
\usepackage{enumitem}
\usepackage{url}
\usepackage{lipsum}

\newcommand*{\origrightarrow}{}

\renewcommand*{\textrightarrow}{\fontfamily{cmr}\selectfont\origrightarrow}

\newcommand{\etal}{\textit{et al.}}

\AtBeginDocument{%
  \providecommand\BibTeX{{%
    \normalfont B\kern-0.5em{\scshape i\kern-0.25em b}\kern-0.8em\TeX}}}

\acmConference[ACM ASIACCS 2023]{ACM ASIA Conference on Computer and Communications Security}{July 10--July 14, 2023}{Melbourne, Australia}

\begin{document}

\title{Mitigating Adversarial Attacks by Distributing Different Copies to Different Buyers}
\author{Jiyi Zhang}
\affiliation{%
  \institution{National University of Singapore}
  \country{Singapore}
}

\author{Han Fang}
\affiliation{%
  \institution{National University of Singapore}
  \country{Singapore}
}

\author{Wesley Joon-Wie Tann}
\affiliation{%
  \institution{National University of Singapore}
  \country{Singapore}
}

\author{Ke Xu}
\affiliation{%
  \institution{Huawei International}
  \country{Singapore}
}

\author{Chengfang Fang}
\affiliation{%
  \institution{Huawei International}
  \country{Singapore}
}

\author{Ee-Chien Chang}
\affiliation{%
  \institution{National University of Singapore}
  \country{Singapore}
}

\begin{abstract}
Machine learning models are vulnerable to adversarial attacks. In this paper, we consider the scenario where a model is distributed to multiple buyers, among which a  malicious buyer attempts to attack another buyer. The malicious buyer probes its copy of the model to search for adversarial samples and then presents the found samples to the victim's copy of the model in order to replicate the attack. We point out that by distributing different copies of the model to different buyers, we can mitigate the attack such that adversarial samples found on one copy would not work on another copy.
We observed that training a model with different randomness indeed mitigates such replication to a certain degree. However, there is no guarantee and retraining is computationally expensive. A number of works extended the retraining method to enhance the differences among models. However, a very limited number of models can be produced using such methods and the computational cost becomes even higher.
Therefore, we propose a flexible parameter rewriting method that directly modifies the model's parameters. 
This method does not require additional training and is able to generate a large number of copies in a more controllable manner, where each copy induces different adversarial regions. 
Experimentation studies show that rewriting can significantly mitigate the attacks while retaining high classification accuracy. For instance, on GTSRB dataset with respect to Hop Skip Jump attack, using attractor-based rewriter can reduce the success rate of replicating the attack to 0.5\% while independently training copies with different randomness can reduce the success rate to 6.5\%. From this study, we believe that there are many further directions worth exploring. 

\end{abstract}

\ccsdesc[500]{Security and privacy~Domain-specific security and privacy architectures}
\ccsdesc[500]{Computing methodologies~Neural networks}

\keywords{security, deep learning, neural networks}

\maketitle

\section{Introduction}
\label{sec:intro}
State-of-the-art machine learning models, such as deep neural networks, are vulnerable to adversarial attacks~\cite{Szegedy2014IntriguingPO}. These attacks generate slightly perturbed samples, which can cause wrong prediction results, even when the perturbed samples retain their original semantics. There are extensive studies conducted on both offenses~\cite{DBLP:conf/kdd/LowdM05,DBLP:conf/iclr/BrendelRB18,Kurakin2017AdversarialEI,Kurakin2017AdversarialML} and defenses~\cite{Shaham2018UnderstandingAT,Parseval_Networks,DBLP:journals/corr/abs-1802-00420} of adversarial attacks. Unfortunately, many attacks are robust and challenging to defend via automated means. There appear to be some intriguing hurdles in the defense against such attacks. 
\\ \\[-3pt]
\noindent
{\em \bf Seller-buyer Distribution Setting.\ \ } 
Instead of the general setting where an attacker attempts to find adversarial samples of a given model, we consider the seller-buyer distribution setting. 
\vspace{-10pt}
\begin{figure}[H]
    \centering 
    \includegraphics[width=0.8\linewidth]{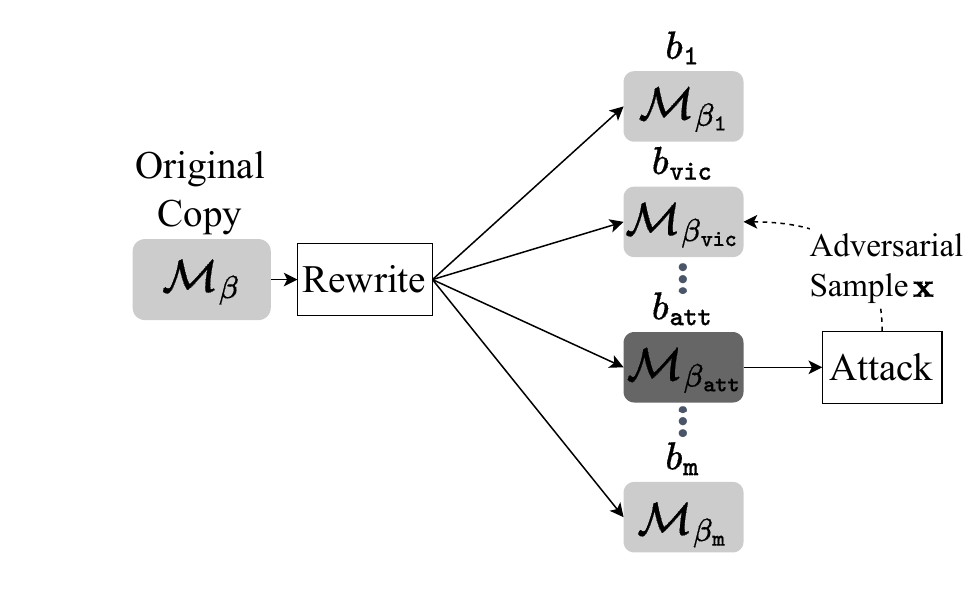}
    \vspace{-15pt}
    \caption{{\em Seller-buyer distribution setting.\ \ } The seller has an original copy  ${\mathcal M}_\beta$.  The seller distributes modified copies ${\mathcal M}_{\beta_i}$ to multiple buyers $b_i$'s. A  malicious buyer $b_{\tt att}$ probes its copy ${\mathcal M}_{\beta_{\tt att}}$ to  obtain an adversarial sample as input for the victim's copy ${\mathcal M}_{\beta_{\tt vic}}$.
    }
    \label{fig:seller-buyer}
\end{figure}
\vspace{-5pt}

Under this setting,  the seller $s$ has a proprietary classification model ${\mathcal M}_\beta$, the {\em original copy},  to be distributed to many subscribed buyers (see Figure~\ref{fig:seller-buyer}).
The seller distributes different copies to different buyers, and each buyer $b_i$ receives a unique copy ${\mathcal M}_{\beta_i}$. The distributed copy can be embedded in physical devices such as autonomous driving vehicles or as a subscribed remote service accessible through a network.  
\newpage
\noindent
{\em \bf Threat Model.\ \ } 
In the seller-buyer distribution setting, we assume that the adversary is a malicious buyer who wants to present an adversarial sample to another victim’s model, and has the following capabilities:\ \
\begin{enumerate}[leftmargin=*]
    \item {\em Access to Own Copy:}\ \ The adversary has black-box access to its respective copy but no direct access to the model parameters.
    \\ \\[-5pt]
    This assumption is plausible in practice:
    \begin{itemize}[leftmargin=*]
        \item In cases where the models are accessed via remote services, the buyers can only send queries and obtain results.
        \item In cases where the models are distributed in physical devices, the adversary can potentially conduct physical attacks to extract the model parameters. In such hostile environment, we can deploy hardware protection mechanism such as trusted execution environments (e.g. Intel SGX\cite{10.1145/2995306.2995307} and ARM TrustZone \cite{trustzone}) to hide the parameters, or embed the parameters into ASIC or FPGA.
    \end{itemize}
    \ \\[-15pt]
    \item {\em No Access to Other User's Copy:}\ \ The adversary does not have intensive black-box access to the victim's copy of the model.
    \\ \\[-5pt]
    Consider autonomous vehicle as an example. In order to generate attack samples, the adversary likely need to feed in a large number of probes (say 10,000 times) to the vehicle. To efficiently probe the vehicle, the adversary would need physical access to the vehicle. For instance, the adversary needs to physically start the vehicle, re-direct the crafted input to the sensors, and observe the outcome. Dismantling of some components might even be required. Under this scenario, the adversary is not able to conduct intensive probing on victim's vehicle.
\end{enumerate}

\noindent
{\em Replication Attack.\ \ }
Since searches of adversarial samples require a large number of probes and the adversary does not have intensive access, the adversary might attempt to probe or attack its own black-box model then presents the crafted input to the victim. For example, an adversary can modify a road sign by probing ML models deployed on its own autonomous vehicle, anticipating the victim passing by the sign.
We call this process replication attack.
\\ \\[-3pt]
\noindent
{\em \bf Mitigating Replication Attack.\ \ } 
The possibility of replication attack leads to the main issue studied in this paper: would adversarial samples generated using an attacker's copy, work on the victim's copy? Clearly, if identical copies are distributed to all users, the adversarial samples found using any copy can always be successfully replicated on every other copy. 

Since the seller could distribute different models to different buyer, we ask this question: Is it possible to rapidly generate many different copies of a model such that the attack is unlikely to replicate across different models?
In other words:
{\em Is it  possible to efficiently generate different copies ${\mathcal M}_{\beta_i}$ from an original copy ${\mathcal M}_{\beta}$, such that adversarial samples found using ${\mathcal M}_{\beta_i}$ are not likely to succeed on copy ${\mathcal M}_{\beta_j}$ when $i\not= j$?  }

If we can adequately answer this question, we would have the means of addressing this security concern in the seller-buyer distribution setting without directly tackling the more challenging task of defending against adversarial attacks in the general setting. 
\\ \\[-3pt]
\noindent
{\em \bf Training-based Solutions to Create Different Models.\ \ } 
One way to obtain different models with similar prediction functionality is by training the models independently on the same dataset but with different randomness. We call this method independent-training. Such randomness is able to induce certain levels of non-transferability among the models and thus mitigate replication attack to some extent. One disadvantage of this method is that the cost of training a new model is expensive. Furthermore, the presence of non-transferability might not be high, and there is no guarantee. Nonetheless, the time complexity is linear and the method could still be practical in some scenarios.

A number of works~\cite{DBLP:journals/access/JalwanaABM20,DBLP:journals/corr/abs-1901-09981,DBLP:conf/ccs/AbdelnabiF21} enhanced the non-transferability by proactively separating the models through training while keeping similar prediction functionality.  For instance, Jalwana~\etal\cite{DBLP:journals/access/JalwanaABM20} proposed a gradient regularization scheme which makes two models orthogonal. Abdelnabi~\etal\cite{DBLP:conf/ccs/AbdelnabiF21} extends adversarial training~\cite{DBLP:journals/corr/GoodfellowSS14} and explicitly penalizes the transferability of attacks when jointly training a set of models together.  Nonetheless, the computation cost is inherently quadratic with respect to the number of models, which is further compounded by the training cost.  Hence, very few models can be produced. For instance, Jalwana~\etal\cite{DBLP:journals/access/JalwanaABM20} and Abdelnabi~\etal\cite{DBLP:conf/ccs/AbdelnabiF21} experimented on $m=2$ and $m=6$ models respectively.  

Perhaps due to the challenge in extending to a larger number of models, Abdelnabi~\etal\cite{DBLP:conf/ccs/AbdelnabiF21} suggests randomly deploying a small pool of $m$ models behind an API to serve buyers so that the probability of adversary accessing the same model as the victim is $1/m$.  However, with the currently achievable $m$, this probability is still too high.
\begin{figure}[H]
    \centering
    \vspace{-5pt}
    \begin{subfigure}{.5\linewidth}
      \centering
      \includegraphics[width=0.95\linewidth]{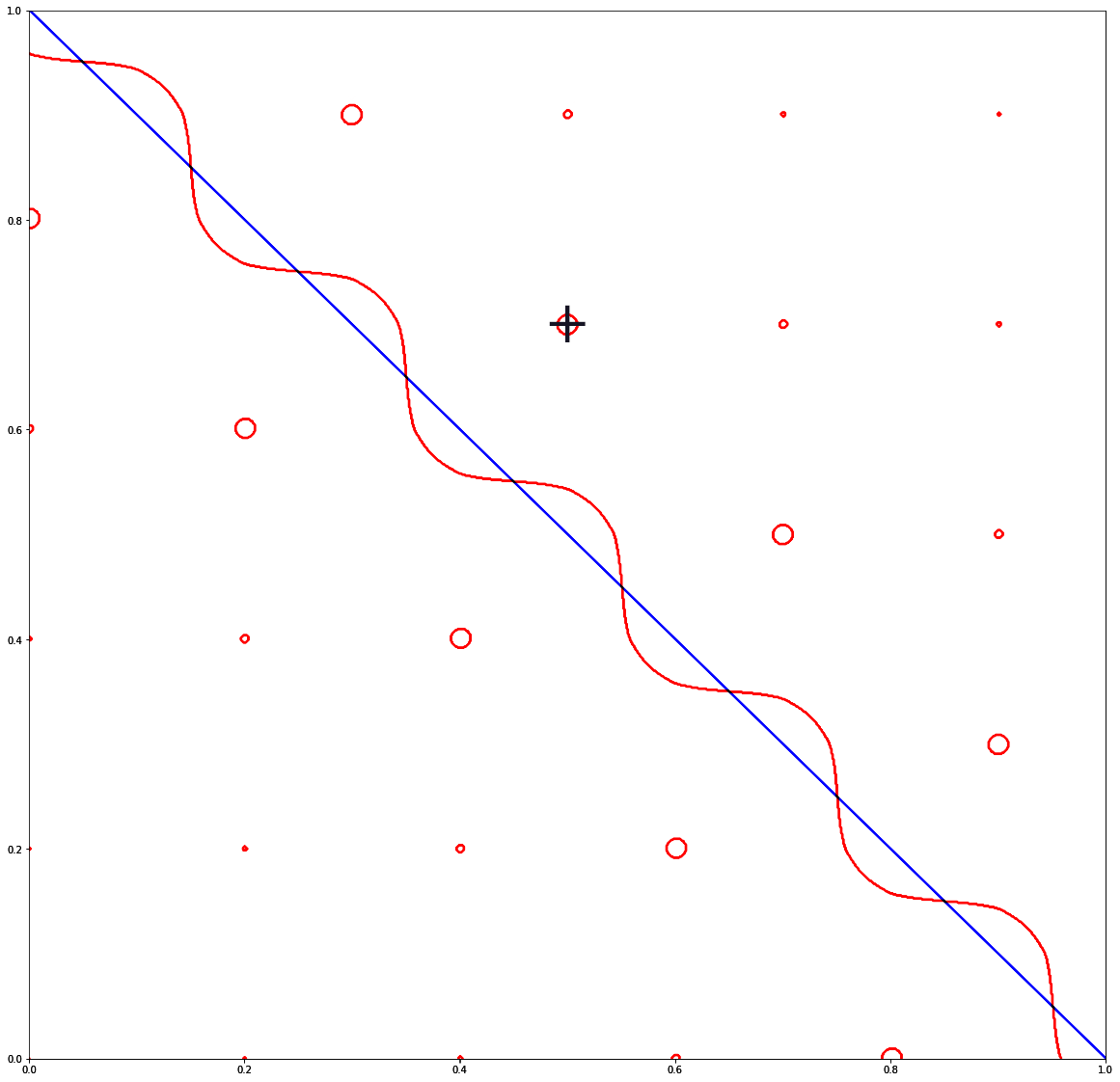}
      \caption{${\mathcal M}_{\beta_{\tt red}}$ and ${\mathcal M}_\beta$.}
      \label{fig:2sub1}
    \end{subfigure}%
    \begin{subfigure}{.5\linewidth}
      \centering
      \includegraphics[width=0.95\linewidth]{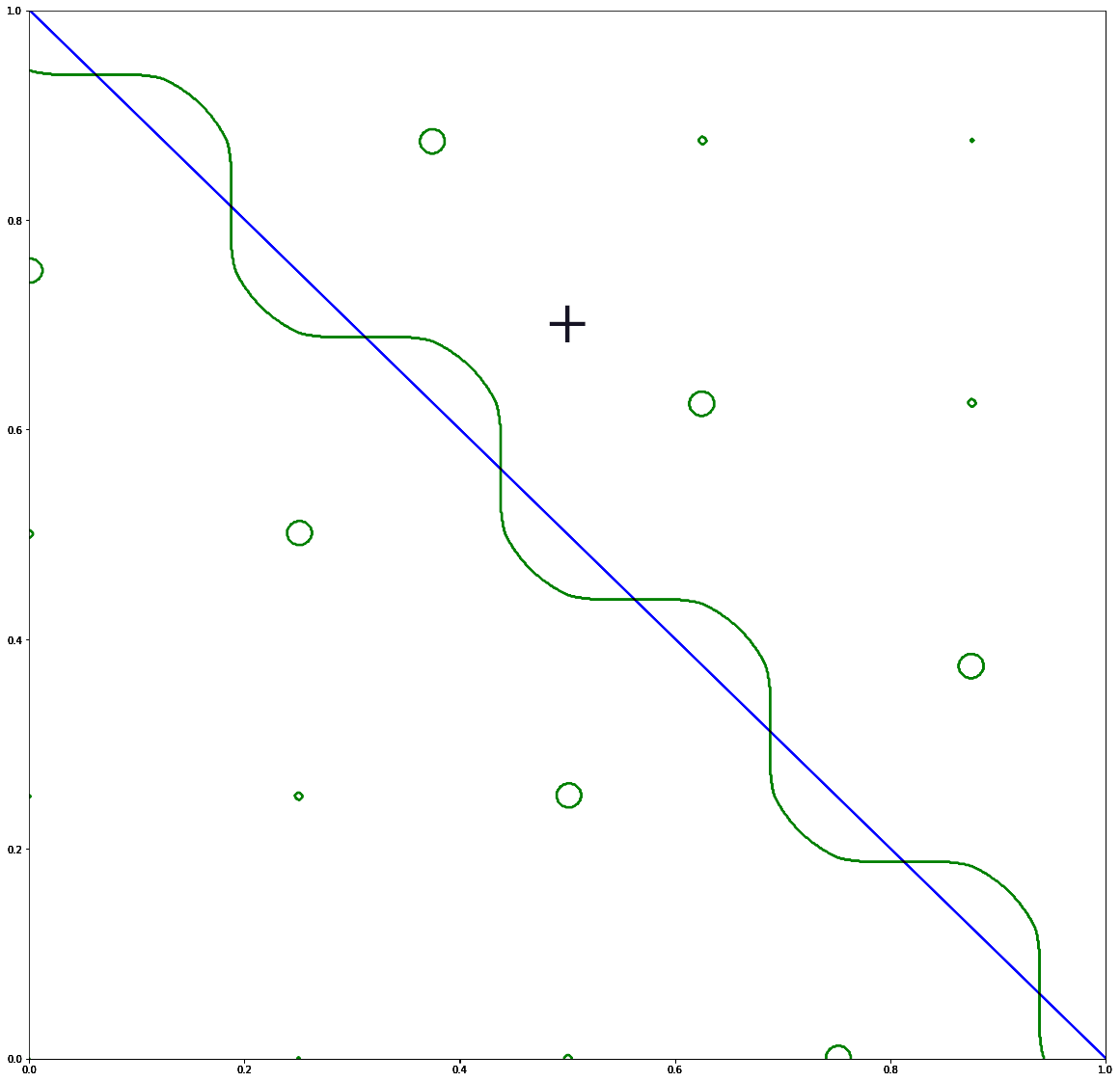}
      \caption{${\mathcal M}_{\beta_{\tt green}}$ and ${\mathcal M}_\beta$.}
      \label{fig:2sub2}
    \end{subfigure}
    \caption{Illustration of decision boundaries in two copies,  ${\mathcal M}_{\beta_{\tt red}}$ and $ {\mathcal M}_{\beta_{\tt green}}$,  of the original copy  ${\mathcal M}_\beta$ when injected with different holes and bumps. Figure (a) and (b) show the  boundary  ${\mathcal M}_\beta$ overlaid with   ${\mathcal M}_{\beta_{\tt red}}$ and ${\mathcal M}_{\beta_{\tt green}}$ respectively. The thick  diagonal straight line indicates the boundary in the original copy
${\mathcal M}_\beta$. 
The symbol ``+'' indicates location of an adversarial example w.r.t.  ${\mathcal M}_{\beta_{\tt red}}$ but is not an adversarial example w.r.t.  ${\mathcal M}_{\beta_{\tt green}}$. }
    \label{fig:2d}
\end{figure}
\vspace{-5pt}
\noindent
{\em \bf Different Adversarial Regions in Different Copies.\ \ }
There are many ways to make copies of a model different. We believe that the most important difference required to mitigate replication attack is the difference in adversarial regions.

We propose {\em parameter rewriting} which induces different adversarial regions in different copies. Parameter rewriting is flexible and modifies the original copy without expensive training. In this paper, we introduce two different types of parameter rewriters. One type is based on attractors~\cite{DBLP:journals/corr/abs-2003-02732} and the other type is based on random permutation. {\em Attractor-based rewriter} explicitly creates new adversarial regions. {\em Permutation-based rewriter} is based on another intuition and binds classes together to modify adversarial regions implicitly. 

Attractor-based rewriter uses an attractor component to attract the attack search paths toward induced adversarial regions. In Figure~\ref{fig:2d}, we give an illustration.
We can visualize the original prediction function ${\mathcal M}_\beta(\bf x)$ in the original copy as a smooth surface. Adversarial samples can be found in regions near the decision boundary.
Attractor-based rewriter creates a new distributed copy by modifying the prediction and scatters steep bumps and holes on the original smooth surface. 
By design, the gradients of the holes and bumps are overwhelmingly steep, which deceive the attackers by guiding the search process toward a nearby hole or bump. Regions near such holes and bumps become new adversarial regions. For any two copies ${\mathcal M}_{\beta_i}$ and ${\mathcal M}_{\beta_j}$, the positions and sizes of holes and bumps are different, controlled by hyper-parameters unique to each copy. We illustrate (see Figure~\ref{fig:2d}) the decision boundaries in two copies of a model where each copy has a different set of holes and bumps. Since the adversarial search process goes to different adversarial regions in different copies, adversarial samples found on one copy unlikely works on another copy. 

Unlike the attractor-based rewriter, permutation-based rewriter is based on another intuition and does not create adversarial regions explicitly. Instead, it creates random bindings internally among the classes of the classifier. For example, suppose we use a permutate ${\mathcal \pi}$ and appoint ${\mathcal \pi}(j)$-th class as the binding `partner' of $j$-th class, the perturbation `effort' to push a sample across the decision boundary for $j$-th class will be split on both the victim $j$-th class and its binding partner ${\mathcal \pi}(j)$-th class. As the bindings are different in different copies, the same amount of `effort' will not be able to push the same sample across the decision boundary in another copy.
\\ \\[-3pt]
\noindent
{\em \bf Construction and Experimentation.\ \ }
We construct three different parameter rewriters. Two of them are based on attractors~\cite{DBLP:journals/corr/abs-2003-02732} and the other one is based on random permutation. 
We investigate replication attacks against these constructions on two datasets CIFAR-10~\cite{Krizhevsky09learningmultiples} and  GTSRB~\cite{Houben-IJCNN-2013}, under three hard-label only black-box attacks, including boundary attack~\cite{DBLP:conf/iclr/BrendelRB18}, Hop Skip Jump attack (HSJ)~\cite{DBLP:conf/sp/ChenJW20} and Geometric Decision-based Attack (GeoDA)~\cite{DBLP:conf/cvpr/RahmatiMFD20}. 
Results show that the proposed method is able to significantly lower the success rate of replication attack at virtually no cost.
We also conduct a preliminary investigation into a {\em collusion attack} where a few malicious buyers collude and combine their copies to search for adversarial samples. Our investigation shows that although the effectiveness of the proposed method decreases as the number of colluders increases, the rate is gradual.
\\ \\[-3pt]
\noindent
{\em \bf Contributions:}
\begin{enumerate}[leftmargin=*]
\item We point out that, in the seller-buyer distribution setting with a large number of buyers, it is possible to readily generate a unique copy for each buyer to mitigate adversarial attacks. 
\item We propose inducing different adversarial regions in different models by a `parameter rewriting' approach. We consider three constructions. Two of them are based on attractors (spread spectrum and QIM) while the third (permutation-based) rewrites by randomly binding different classes.
Experiments on three different attacks show that the proposed method effectively diminishes the success rate of replication attacks among models.
\item We conducted a preliminary investigation on the collusion attack. Our analysis shows that although the effectiveness of defense decreases gradually as the number of colluders increases, the method is still adequate for a reasonably large number of colluders.
\end{enumerate}
\section{Related Works}
\label{sec:related}
\subsection{Adversarial Attacks}
Adversarial attacks aim to find small distortion on input samples that leads to the wrong prediction results, and these attacks can be formulated as an optimization problem. To the best of our knowledge, all known efficient attacks utilize some objective functions to guide the search processes of the adversary and can be categorized into gradient-based attacks and non-gradient-based attacks.
\\ \\[-3pt]
\noindent
{\em \bf Gradient-based Attacks. \ \ }
Many white-box attacks use the backpropagated gradients. During normal training, for a given input, the pair of actual output and the expected output is fed into the loss function to compute a loss. The backpropagated gradients are used to update the model's parameters. During an attack, the attacker fixes the model's parameters and uses an attack-objective function to compute the loss between the model's actual output and the adversary's desired output. The backpropagated gradients are then used to update the input and make it adversarial.  
The attack's objective function can be either derived directly from the training loss function~\cite{Goodfellow2015ExplainingAH,Kurakin2017AdversarialEI,Madry2017TowardsDL,DBLP:journals/corr/abs-1710-06081}, optimized with less distortion~\cite{MoosaviDezfooli2016DeepFoolAS,Szegedy2014IntriguingPO,towards,chen2018ead}, or optimized with special goals~\cite{DBLP:journals/corr/Moosavi-Dezfooli16,Papernot2015TheLO}. 

One of the attack methods, the Fast Gradient Sign Method (FGSM), is a simple and
fast approach~\cite{Goodfellow2015ExplainingAH}. It moves a fixed small step $\epsilon$ in the direction that maximally changes the prediction result. The adversarial sample is: $$
{\textbf x}'={\textbf x}+\epsilon \cdot \operatorname{sign}\left(\nabla_{{\textbf x}} J\left(\theta, {\textbf x}, {\textbf y}_{true}\right)\right)
$$ where ${\textbf y}_{true}$ is the one-hot vector of the true label of the input ${\textbf x}$. PGD~\cite{Madry2017TowardsDL}, extends the FGSM from a one-step attack into an iterative process. The search process starts at a random point within the norm ball. The DeepFool~\cite{MoosaviDezfooli2016DeepFoolAS} attack views neural network classifiers as hyperplanes separating different classes and finds the minimal perturbation $\epsilon$ to change the prediction result.
Last but not least, the C\&W~\cite{towards} attack is an iterative optimization method. Its goal is to minimize the loss $\left| \epsilon\right| + c \cdot f({\textbf x}+\epsilon)$ where $f$ is an objective function such that the adversarial sample ${\textbf x}+\epsilon$ is classified into the target class if and only if $f({\textbf x}+\epsilon) \leq 0$ and $c$ is a constant. 
\\ \\[-3pt]
\noindent
{\em \bf Non-gradient Based Attacks.\ \ }
Black-box attacks cannot rely on backpropagation. Instead, they use some form of sampling and optimization techniques. Their mechanisms can be divided into score-based~\cite{DBLP:conf/iclr/RuCBG20,DBLP:journals/corr/abs-1910-02244}, decision-based~\cite{DBLP:conf/iclr/BrendelRB18,Cheng2020Sign-OPT}, and gradients approximation~\cite{DBLP:journals/corr/abs-1802-00420,DBLP:journals/corr/abs-1802-05666}. A decision-based attack, the boundary attack~\cite{DBLP:conf/iclr/BrendelRB18}, is an attack that starts from a large adversarial perturbation and then seeks to reduce the perturbation while staying adversarial. 

\subsection{Transfer Attacks}
Adversarial samples can be transferable across models with different architectures and trained with similar datasets. In particular, when attacking a substitute model cloned from a victim~\cite{Papernot:2017:PBA:3052973.3053009}, some of the generated adversarial samples can also cause misclassification on the original victim model. 
Therefore, an attacker can apply model extraction attacks~\cite{DBLP:conf/uss/TramerZJRR16} on a black-box victim, conduct known white-box attacks on the extracted model, and then collect adversarial samples that attack the original black-box victim. Transfer-based adversarial attacks in image classification models often achieve limited success due to the overfitting of local models. However, mechanisms~\cite{Wu_2020_CVPR,Wu_2021_CVPR} have been proposed to circumvent the overfitting issue, promoting the transferability of adversarial samples. 
While empirical evidence shows that transfer attacks are successful, Demontis~\etal~take it a step further and perform a comprehensive analysis~\cite{236234} to study underlying reasons for the transferability of attacks.

\subsection{Attractors}
Most attacks search for adversarial samples along directions derived from some local properties. For instance, FGSM~\cite{Goodfellow2015ExplainingAH} derives the search direction from the training loss function's gradient. 
The attractors are from a proposed method~\cite{DBLP:journals/corr/abs-2003-02732} that detects adversarial perturbations, which considers characteristics of the attack process. 
This method views each attack as an optimization problem. The goal of the attractors, injected into the classifier model, is to mislead the adversary by actively modifying the optimization objectives. An attractor ${\textbf x}_0$ effectively adds artifacts that taint the local properties of a search space, such that the gradients of the attack objective function point towards that attractor. 

The attractors serve two purposes. First, they introduce cavities into an otherwise smooth search space, which confuses an adversary's search process. Next, they attract the search process to some dedicated regions in the space, leading adversarial samples into an easily detected position.

For a classification model $\mathcal{M_\theta}$ parameterized by $\theta$, a robust digital watermarking scheme is first chosen, along with a watermark decoder model $\mathcal{W_\phi}$ that is parameterized by $\phi$. Next, the classification model  $\mathcal{M_\theta}$ and the decoder model $\mathcal{W_\phi}$ are combined to form a new model $\mathcal{M_{\theta, \phi}}$. Given an input sample ${\textbf x}$, this new model $\mathcal{M_{\theta, \phi}}$ returns the normalized sum of both the classifier and decoder model outputs $(\mathcal{M_\theta}(\textbf x) + \mathcal{W_\phi}(\textbf x))$, which is available to an adversary. Hence, the addition of this watermark makes an attack unintentionally affect the correlation of the corresponding watermark, allowing for the detection of adversarial samples.

The above properties of the attractor made it an appropriate candidate for the parameter rewriting. One of the rewriting methods we proposed uses attractor-based rewriter. It is modified based on the original attractor with two major differences:
\begin{itemize}[leftmargin=*]
    \item {\em Emphasis on Attracting vs. Detection.}\ \ The original attractor is designed to guide the attacking process into a position that could be easily detected. This detection requirement is not applicable in the seller-buyer distribution setting. The difference in the goals leads to different choices of parameters.
    \\ \\[-10pt]
    \item {\em Different Attractors for Different Buyers.}\ \ The original attractor considers the defense of a single model. As our setting involves multiple copies of a model, different attractors have to be injected into different copies. The same attack process is attracted to different local minima in different copies, and samples generated on one copy are not adversarial in another copy. A good attractor candidate may not work in our case if we cannot efficiently produce sufficiently different variants of it to accommodate a large number of copies.
\end{itemize}
\section{Problem Formulation}

\subsection{Seller-buyer Distribution Setting}

In the seller-buyer distribution setting, a seller $s$ has  the original copy of a classification model, which is to be distributed to multiple buyers $b_1, b_2, \ldots, b_m$, and  each buyer might receive a different copy.   Each buyer  $b_i$ has an identity, and for simplicity, we take index $i$ as its identity.  For simplicity, we restrict the type of classification models to be neural networks based classifiers.  Below are the details.
\\ \\
\noindent
{\em \bf Seller and Original Copy.}\ \
The seller has a training dataset ${\mathcal T}$, from which the seller  trains a classification model  ${\mathcal M}_\beta$, which we call the {\em original copy}.   The subscript $\beta$,  also known as the model parameters, encodes   information of the neural network's weights and architecture and possibly other information required to compute ${\mathcal M}_\beta$.  Hence, $\beta$ can be treated as a  computer program that computes the function.
The input of $ {\mathcal M}_\beta $ is a sample from ${\mathcal X}$, and the output is the corresponding prediction.  The training process  could be probabilistic, using different seeds as the  source of randomness.   For instance, the training process might update the model using randomly selected training samples, with randomness  derived from the seed.  The  seed for the original copy is not crucial. Nonetheless, let us denote it as $k_0$.
\\ \\
\noindent
{\em \bf Distributed Copies.}\ \
For each buyer $b_i$, the seller generates a machine learning model ${\mathcal M}_{\beta_i}$. 
The generation methods can be broadly classified into two types based on whether the training dataset is involved in the generation process. If a method utilizes the training dataset, we categorize it as a training-based method. Otherwise, we classify it as a parameter rewriting method. 
Training-based methods could be expensive due to two factors. Firstly, some methods~\cite{DBLP:journals/access/JalwanaABM20,DBLP:journals/corr/abs-1901-09981,DBLP:conf/ccs/AbdelnabiF21} conduct joint training and thus the cost grows significantly (e.g. quadratic) as the number of models increases. Secondly, training for a single model is also becoming very expensive due to the growing size of model architectures and datasets.
There is a straightforward training-based method which we term as ``independent-training'', where $m$ models are trained independently with different randomness. Independent-training is still expensive due to the above second factor. Nonetheless, it is linear with respect to the number of copies and could still be feasible in some scenarios. 
\begin{itemize}[leftmargin=*,topsep=5pt]
  \item {\em Independent-training} trains ${\mathcal M}_{\beta_i}$ from the training dataset ${\mathcal T}$ using different seeds.   More specifically, the method takes  $\beta$,  $k_i$, and ${\mathcal T}$ as inputs, where $k_i$ is the  seed.  The seed $k_i$ can be extracted from a truly random source  or  generated from $(i, k)$  using a  secure pseudo-random number generator.  For each buyer $b_i$, the method carries out   training  on dataset ${\mathcal T}$  with  $k_i$ as the seed, and obtains ${\mathcal M}_{\beta_i}$. In our evaluation, it does not utilize the parameter $\beta$  and simply applies the same training procedure for all buyers.
  \vspace{10pt}
  \item {\em Parameter rewriting} takes in $\beta$ and $k_i$ as inputs, generating the  model ${\mathcal M}_{\beta_i}$, where $k_i$ is some seed.   Likewise, the seed $k_i$ can be extracted from some truly random sources or generated from $(i,k)$ using some secure pseudo-random number generator. 
  Unlike the training-based method,  rewriting modifies  the weights and architecture of the original copy  ${\mathcal M}_\beta$ without using the training dataset ${\mathcal T}$, in analogy to program rewriting where programs are being modified.   In addition, the modification is parameterized by $k_i$, and thus ${\mathcal M}_{\beta_i}$ could be different for different buyers.
\end{itemize}
The $k$ and all $k_i$'s must  be kept secret from the buyers. The seller does not need them subsequently, and thus they can be discarded after the distributed copies have been generated.  
It is possible to apply training and rewriting together.  For instance, the seller can independently train two copies ${\mathcal M}_{\phi}$ and ${\mathcal M}_{\psi}$. Next, for two different buyers, $b_i$ and $b_{j}$, the seller  rewrites ${\mathcal M}_{\phi}$ to obtain ${\mathcal M}_{\phi_i}$ and 
rewrites ${\mathcal M}_\psi$ to obtain ${\mathcal M}_{\psi_{j}}$.    Finally, 
${\mathcal M}_{\phi_i} $ and ${\mathcal M}_{\psi_{j}}$ can be distributed to $b_i$ and $b_{j}$, respectively. 
\\ \\
\noindent
{\em \bf Black-box and Hard-label Access.}\ \
Each buyer $b_i$ has black-box access to its model ${\mathcal M}_{\beta_i}$.
That is,  $b_i$ can adaptively submit any sample  $\textbf x$ and obtain the hard prediction $\argmax({\mathcal M}_{\beta_i} (\textbf x))$.  However, $b_i$ cannot directly obtain the model parameters of  ${\mathcal M}_{\beta_i}$.

In practice, black-box access can be realized by remote access, where  each buyer logs in to the seller's server, sends  the samples to the server, and receives the hard predictions from the server.   Alternatively,  the model ${\mathcal M}_{\beta_i}$ can be embedded into physical devices with ASIC or FPGA, or Trusted Execution Environment  (TEE)  such as Intel SGX\cite{10.1145/2995306.2995307} and ARM TrustZone \cite{trustzone}.  The devices  are to be distributed to the buyers  and placed under the buyers' control.  The buyers can invoke the classifier, send in samples and obtain the hard predictions,  but are unable to extract the model parameters from the tamper-proof hardware.

While TEE  relies on hardware protection, it would be useful to have a cryptographic solution that facilitates black-box access while hiding the parameters. 
This can be achieved by functional encryption, which, unfortunately, does not have an efficient solution yet. However, recent breakthroughs~\cite{DBLP:conf/stoc/JainLS21,DBLP:journals/siamcomp/GargGH0SW16} could lead to practical implementation. 

For both remote access and TEE,  it is possible to enforce access control so that each buyer  is unable to submit an extremely large number of samples. As the number of accesses depends on applications, in our evaluation, we treat the number of black-box accesses in the order of $10^6$ as feasible. 

\subsection{Threat Model}
A malicious buyer $b_{\tt att}$ intends to compromise a victim buyer $b_{\tt vic}$ on a  given sample $\textbf x$. 
Given   $\textbf x$, the attack is successful if $b_{\tt att}$ finds a sample $\textbf x'$ with respect to   ${\mathcal M}_{\beta_{\tt vic}}$,  that is, the difference between $\textbf x'$ and $\textbf x$ is small, but yet the classes predicted by ${\mathcal M}_{\beta_{\tt vic}}$ are different for $\textbf x'$ and $\textbf x$. Specifically:
\begin{equation}
\label{eq:argmax}
\argmax {\mathcal M}_{\beta_{\tt vic}} (\textbf x) \not =\argmax {\mathcal M}_{\beta_{\tt vic}} (\textbf x').     
\end{equation}
\noindent
{\em \bf Capability of Malicious Buyer.}\ \
The malicious buyer $b_{\tt att}$ has black-box and hard-label access to its copy ${\mathcal M}_{\beta_{\tt att}}$.    It might know the training procedure and the algorithm in generating the distributed copies,  but we assume that $b_{\tt att}$ does not know the training dataset ${\mathcal T}$ and the seeds that generate the original and distributed copies.   

We assume that the malicious buyer does not have extensive access to the victim's classifier. Specifically, this malicious buyer has only one or a few opportunities to present input samples to the victim ${\mathcal M}_{\beta_{\tt vic}}$. 
\\ \\[-3pt]
\noindent
{\em \bf Targeted vs. Un-targeted attack.}\ \
As indicated in Equation~\ref{eq:argmax}, the attacker's goal is to find a sample that is being misclassified to any class other than the correct class. This is known as an {\em un-targeted attack}. Alternatively, the attacker could have a more restricted goal of a {\em targeted attack}. In such cases, when given a sample ${\textbf x}$ and a particular class, the malicious buyer may have a  specific goal of finding an adversarial sample misclassified as the particular class. Clearly, to an adversary, the un-targeted attack is a weaker goal and is easier to achieve. 
\\ \\[-3pt]
\noindent
{\em \bf Collusion Attack.}\ \
A single attacker might purchase or subscribe to the model under multiple names and have black-box access to multiple copies. We call this a {\em collusion attack}, as it is equivalent to a few colluding buyers who combine their models. Intuitively, access to multiple black-boxes could aid the malicious buyers in improving the replication process. 

Since our framework is new, currently, there is no known collusion attack. For this preliminary investigation, we designed a custom collusion attack. Our experiments show that while the attack success rate increases with more colluders as expected, the rate of increase is gradual (see Section~\ref{sec:analysis}).

This attack is similar to collusion attacks in digital watermarking, where the attacker attempts to remove watermarks by combining multiple watermarked copies of the same host~\cite{DBLP:journals/tip/WangWZTL05}.
An approach in mitigating watermark collusion attacks is via anti-collusion fingerprinting code~\cite{DBLP:journals/tit/SkoricVCT08}. It would be interesting to investigate this application further.

\subsection{Relationship to Transfer Attack}
Replicating an attack in the seller-buyer distribution setting can be considered as a special case of transfer attack. As every copy of the model shares the same architecture and training data, replicating an attack is much easier to achieve when compared with a normal transfer attack where different architectures and datasets are involved. 
\section{Proposed Method: Parameter Rewriting}
\subsection{Main Idea}
\label{sec:main_idea}
Given the original copy ${\mathcal M}_\beta$ and a random seed $k_i$ for some buyer $i$, we want to rewrite $\beta$ and obtain another copy
${\mathcal M}_{\beta_i}$ without using the training dataset.

In this paper, parameter rewriting is achieved through modifying the original copy ${\mathcal M}_\beta$ with a rewriter function ${\mathcal R}_{k_i}$. The rewriter ${\mathcal R}_{k_i}$ can be an explicit extra component or an implicit processing step which modifies the output.  

Recap that the seed $k_i$ is to be kept secret from the buyers, and can be treated as the secret key.  When ${\mathcal R}_{k_i}$ is in the form of an extra component, it is not necessary that ${\mathcal R}_{k_i}$ is encoded as  neural network models. 
For instance, it can be implemented using some signal processing libraries.  Nonetheless, in our evaluation, we implement them using given framework (PyTorch) to facilitate evaluation on known attacks.

\subsection{Considerations for ${\mathcal R}_{k_i}$}
The process of generating an adversarial sample can be seen as a search process where the attacker starts from the input sample and crosses a nearby decision boundary so that  the sample would be misclassified.  To find the nearest adversarial sample, optimization techniques are often employed to minimize the distance between an adversarial sample and its starting point. Most adversarial attacks, either explicitly or implicitly,  utilize an attack objective function for such optimization. Clearly, when attacking two identical copies of a model,  the  search paths to  the adversarial sample would be the same. The role of the rewriter ${\mathcal R}_{k_i}$ is to separate them.

Intuitively, the rewriter should  have the following properties:  (1) it induces different adversarial regions with different $k_i$'s; (2) it does not to degrade classification accuracy.
\\ \\[-3pt]
We consider two types of parameter rewriters: explicit attractor-based~\cite{DBLP:journals/corr/abs-2003-02732} rewriter and implicit permutation-based rewriter. They modify adversarial regions based on different mechanisms. We give the details of their mechanisms in Section~\ref{sec:attractor_rewriter} and~\ref{sec:permutation_rewriter}. In total, we constructed three candidates:  
\begin{enumerate}[leftmargin=*]
    \item Explicit attractor-based rewriter using spread spectrum decoder.
    \item Explicit attractor-based rewriter using Quantization Index Modulation (QIM) decoder.
    \item Implicit permutation-based rewriter.
\end{enumerate}
\  \\[-10pt]
\noindent
{\em \bf Remarks. \ \ } The first candidate is mainly used for analysis in Section~\ref{sec:analysis} to show the influence on local gradient. As its performance is not ideal, we recommend using the second candidate (attractor-based rewriter with QIM decoder) and the third candidate (permutation-based rewriter) for defense against actual attacks and we show the results of evaluation in Section~\ref{sec:evaluation_under_attack}.

For attractor-based rewriters, our implementation of attractor is slightly different from the attractors used in direct defense~\cite{DBLP:journals/corr/abs-2003-02732}. The original attractor is designed to guide the attacking process into a position that could be easily detected. Since detection is not applicable in the seller-buyer distribution setting, we use a modified version of attractors. We do not use hard constraints to force an adversarial sample to move near to an attractor as we no longer need to maximally differentiate adversarial samples from clean samples and make them detectable.

\subsection{Attractor-based Rewriter}
\label{sec:attractor_rewriter}
The attractor~\cite{DBLP:journals/corr/abs-2003-02732} was originally designed as a direct defense which detects adversarial perturbations by evaluating characteristics of the attack process. Attractors, when injected into a model, modify the local properties and attract attack objective function to some preset zones. These preset zones can be viewed as new adversarial regions created by attractors. Based on this mechanism, a parameter rewriter can inject different attractors to different copies to induce different adversarial regions. Therefore, adversarial samples found in adversarial region from one copy will not be able to attack another copy with different adversarial region.
\\ \\[-3pt]
\noindent
{\em \bf Construction.} \ \
We adopt the `summation'' method: the generated model  ${\mathcal M}_{\beta_i}$  is pieced together using the original copy and the attractor-based rewriter.
Specifically, on input ${\textbf x}$, the  output is the normalized (w.r.t. L1 norm) sum:
\begin{equation}
\label{eq:nsum}
{\mathcal M}_{\beta_{i}}({\textbf x}) = \frac{ {\mathcal M_{\beta}} ({\textbf x}) + {\mathcal R}_ {k_{i}}({\textbf x}) } { \| {\mathcal M_{\beta}} ({\textbf x}) + {\mathcal R}_{k_{i}}({\textbf x})   \|_1}
\end{equation}
Different ${\mathcal R}_{k_{i}}$ are used for each buyer $u_i$. The seed $k_i$ is used to generate watermarks for each of the $n$ classes. For each class, the watermark can be in the form of a single message or a set of messages. Suppose we apply single message and select  $n$ messages,  ${\textbf m}_j$,  for $j=1, \ldots, n$. By the summation  in equation~\ref{eq:nsum},  the message ${\textbf m}_j$ is tied to the $j$-th class of the classification model ${\mathcal M}_{\beta_i}$   for any $j$.  Any attempt to increase/decrease the prediction score of  the $j$-th class in  ${\mathcal M}_{\beta_i} ({\textbf x})$ would unknowingly increase/decrease the correlation with the message ${\textbf m}_j$ in $\mathcal{R}_{k_i} ({\textbf x})$.

When ${\mathcal R}_{k_{i}}$ is dominant, most of the perturbation caused by the attack will be contributed by ${\mathcal R}_{k_{i}}$. Since different ${\mathcal R}_{k_{i}}$ are used in different copies, when replicating an adversarial sample, the dominant part will no longer increase/decrease in adversary's intended manner, thus making the attack ineffective. Here we consider two candidates: spread spectrum decoder and QIM decoder.
\\ \\[-3pt]
\noindent
{\em \bf Spread Spectrum Decoder as ${\mathcal R}_{k_i}$.}\ \
For each buyer $u_i$,   we  randomly (using $k_i$ as source of randomness) select  $n$ messages,  ${\textbf m}_j$,  for $j=1, \ldots, n$, where each ${\textbf m}_j$ is a vector with  $\ell$ real coefficients,  and $\ell$ is the number of pixels in a sample.  

On input ${\textbf x}$, 
$\mathcal{R}_{k_i} ({\textbf x}) = \langle {c}_1, c_2, \ldots, c_n\rangle, $ 
where $c_j =  A \ {\textbf x} \cdot {\textbf m}_j $ for each $j$,   $A$ is a pre-defined  constant  that is the same for all buyers,  and the operation $\cdot$ is the inner product
which essentially transform the pixel values to a feature space.
Hence, each $c_j$ gives the correlation with the $j$-th message. 
\\ \\[-3pt]
\noindent
{\em \bf QIM Decoder as ${\mathcal R}_{k_i}$.}\ \
For each buyer $b_i$,  a set of messages ${\textbf m}_{j,h}$ are randomly (using $k_i$ as source of randomness) selected, for $j=1, \ldots, n$, $h=1, \ldots, B $,  where each ${\textbf m}_{j,h}$ is a vector with  $\ell$ real coefficients and  $B$ is some pre-defined constant. Other predefined parameters are the step size $\delta$,  and a set of weights $\alpha_j$ for $j=1, \dots, B$,  which are the same for all buyers.   Here, the set $\langle {\textbf m}_{j,1}, \ldots {\textbf m}_{j,B} \rangle$ correspond to the $j$-th watermark.

 On input ${\textbf x}$,  for each $j=1, \ldots n$  the following steps are carried out:
 \begin{enumerate}
 \item Compute $y_h = {\textbf m}_{j,h} \cdot {\textbf x}$ for $h=1, \ldots, B$.   That is, the sample ${\textbf x}$ is projected to a $B$-dimensional vector $(y_1, \ldots, y_B)$.
\item Compute  the total weighted quantization error,
          $$  e'_j=   \sum_{j=1}^B   \alpha_j  |  y_j +\delta/2- round(y_j/\delta+1/2)\times \delta  |$$
\end{enumerate}
 where $round(\cdot)$ returns the nearest integer of input value.
 
 The value $e'_j$ is the distance of the sample ${\textbf x }$ from the $j$-th watermark.  We further linearly map it to the  range of $[0,1]$  and larger distances are mapped to smaller values.  Let $e_j$ be the mapped value.  Overall,  on input ${\textbf x}$,  we have
$\mathcal{R}_{k_i} ({\textbf x}) = \langle e_1, \ldots, e_n \rangle$.

\subsection{Permutation-based Rewriter}
\label{sec:permutation_rewriter}
Attractor-based rewriters are able to rewrite the parameters without additional training. However, they still add an extra component to the original copy. Here we propose a permutation-based rewriter without any extra component as an alternative. Instead of binding additional watermark messages to each class, we create random bindings internally among the classes of the classifier.
\\ \\[-3pt]
\noindent
{\em \bf Construction.} \ \
We set ${\mathcal R}_ {k_{i}}({\textbf x}) = \alpha \cdot {\mathcal \pi}_ {k_{i}}({\mathcal M_{\beta}} ({\textbf x}))$.
On input ${\textbf x}$, the output is the normalized (w.r.t. L1 norm) sum:
\begin{equation}
    {\mathcal M}_{\beta_{i}}({\textbf x}) = \frac{ {\mathcal M_{\beta}} ({\textbf x}) + \alpha \cdot {\mathcal \pi}_ {k_{i}}({\mathcal M_{\beta}} ({\textbf x})) } { \| {\mathcal M_{\beta}} ({\textbf x}) + \alpha \cdot {\mathcal \pi}_{k_{i}}({\mathcal M_{\beta}} ({\textbf x}))   \|_1}
\end{equation}
\noindent
where ${\mathcal \pi}_{k_{i}}$ is a random permutation matrix with $k_{i}$ as the random seed. For the convenience of notation, suppose $j$ is the original class, we also use ${\mathcal \pi}_{k_{i}}(j)$ to denote the new class after the permutation.

To see why this approach works, note that any attempt to increase/decrease the prediction score of the $j$-th class in  ${\mathcal M}_{\beta_j} ({\textbf x})$ will need to increase/decrease the score for both $j$-th class and ${\mathcal \pi}_{k_{i}}(j)$-th class. The amount of change in $j$-th class and ${\mathcal \pi}_{k_{i}}(j)$-th class depends on the choice of a predefined parameter $\alpha$. By using a large $\alpha$, a significant amount of increase/decrease needs to be from ${\mathcal \pi}_{k_{i}}(j)$-th class to push the perturbed sample across decision boundary. However, as ${\mathcal \pi}_{k_{i}}(j)$ is different in different copies, the perturbation which is able to successfully attack a sample for one copy will not be able to push it cross the decision boundary in another copy. 

Our evaluation shows that we can indeed choose a very large $\alpha$ (0.8 for CIFAR10 dataset) without affecting classification accuracy. Samples from testing datasets show that the difference between the highest prediction score and the second highest prediction score (after softmax) is almost always very large on any clean natural input.
\\ \\[-3pt]
\noindent
{\em \bf Remarks.}\ \
Unlike attractor-based rewriters, the number of copies that can be generated using permutation-based rewriter is associated with the number of classes in the original copy. For an $n$-class classifier model, the number of copies is limited to $n!$. To deal with models with few classes, special techniques may be required. For example, one workaround is to add dummy classes then permutate.
\section{Evaluation}
\subsection{Setup}
\noindent
{\em \bf Dataset.}\ \
Our evaluation was conducted on two different datasets: CIFAR-10~\cite{Krizhevsky09learningmultiples} (Canadian Institute For Advanced Research)
and GTSRB~\cite{Houben-IJCNN-2013} (German Traffic Sign Recognition Dataset). CIFAR-10 contains 10 classes of animals and vehicles, split into 50,000
training images and 10,000 testing images. GTSRB contains 43 classes of traffic signs, split into 39,209 training images and 12,630 testing images. We resize all the images to $32 \times 32 \times 3$.
\\ \\[-3pt]
\noindent
{\em \bf Model.}\ \
The evaluations are conducted for three different rewriters: attractor-based rewriter using spread spectrum decoder, attractor-based rewriter using QIM decoder and permutation-based rewriter. 
\\ \\
We first constructed two models using the independent-training method as references.
\begin{itemize}[topsep=5pt]
    \item $\mathcal{M}_{\phi}$: Classifier trained with seed $p$ and parametrized by $\phi$.
    \item $\mathcal{M}_{\psi}$: Classifier trained with seed $q$ and parametrized by $\psi$.
\end{itemize}
\noindent
For each rewriter, we constructed three models on top of the reference models for the experiments:
\begin{itemize}[topsep=5pt]
    \item $\mathcal{M}_{\phi_1}$: $\mathcal{M}_{\phi}$ using rewriter ${\mathcal R}_{k_1}$ and seed $k_1$.
    \item $\mathcal{M}_{\phi_2}$: $\mathcal{M}_{\phi}$ using rewriter ${\mathcal R}_{k_2}$ and seed $k_2$.
    \item $\mathcal{M}_{\psi_2}$: $\mathcal{M}_{\psi}$ using rewriter ${\mathcal R}_{k_2}$ and seed $k_2$.
\end{itemize}

We constructed these models for both MNIST and GTSRB datasets. We used VGG-19~\cite{DBLP:journals/corr/SimonyanZ14a} as the base architecture and trained each model for 200 epochs. For $\mathcal{M}_{\phi}$ and $\mathcal{M}_{\psi}$, the seeds determine the iteration order of dataloader and parameters of data augmentation such as resized crop for each data point. For $\mathcal{M}_{\phi_1}$, $\mathcal{M}_{\phi_2}$ and $\mathcal{M}_{\psi_2}$, the seeds determine the watermark messages embedded in the decoders (attractor-based rewriter) or the order of permutation (permutation-based rewriter).

\subsection{Accuracy on Testing Datasets}
The rewriters modify the outputs of the copies and thus could decrease the accuracy.
We evaluated the accuracy of models obtained through parameter rewriting on testing datasets (10,000 testing images for CIFAR-10 and 12,630 testing images for GTSRB) and reported the results in Table \ref{tab:onclean}. In this experiment, we use model ${\mathcal M}_{\phi}$ and model ${\mathcal M}_{\phi_1}$. 

\vspace{-5pt}
\begin{table}[H]
    \centering
    \small
    \begin{tabular}{|c|c|c|} 
        \hline
        Datasets                                                                            & CIFAR-10          & GTSRB              \\ 
        \hline
        Original Copy                                                                       & 96.3\% (1.3\%)  & 95.8\% (1.7\%)   \\ 
        \hline
        \begin{tabular}[c]{@{}c@{}}Attractor-based Rewriter\\(Spread Spectrum)\end{tabular} & 94.3\% (0.3\%) & 95.1\% (0.3\%)  \\ 
        \hline
        \begin{tabular}[c]{@{}c@{}}Attractor-based Rewriter (QIM)\end{tabular}             & 94.5\% (0.2\%)  & 95.3\% (0.3\%)   \\ 
        \hline
        Permutation-based Rewriter                                                          & 94.0\% (1.0\%)  & 95.1\% (0.8\%)   \\
        \hline
        \end{tabular}
    \caption{Accuracy of original model and new copies generated by parameter rewriters. The first value is the average accuracy, and the second value in the bracket is the standard deviation.}
    \label{tab:onclean}
\end{table}
\vspace{-15pt}

For each evaluation, we create 10 copies of the model and compute the average accuracy. If the model is obtained through training (e.g. the original copy), we train 10 copies from scratch. If the model is obtained through controlling seeds, we use different random seeds to generate 10 copies.
Our results show that the classification accuracy of new copies generated by attractor-based rewriter and permutation-based rewriter are both very close to the original copy.

\subsection{Influence on Local Gradient (Attractor-based Rewriter)}
\label{sec:analysis}
Based on attractor-based rewriters' mechanism of inducing new adversarial regions, we can check their effectiveness first without conducting attacks. This analysis is done by inspecting rewriter's influence on local gradient.

In this section, we conduct an experiment using the attractor-based rewriter with spread spectrum decoder and attractor-based rewriter with QIM decoder. For each rewriter, we use the 10,000 testing images from CIFAR-10. For each image, we compute the cosine similarity of the gradients with regard to the loss function in the following three pairs:
\begin{itemize}
    \item $\mathcal{M}_{\phi}$ and $\mathcal{M}_{\psi}$
    \item $\mathcal{M}_{\phi_1}$ and $\mathcal{M}_{\phi_2}$
    \item $\mathcal{M}_{\phi_1}$ and $\mathcal{M}_{\psi_2}$
\end{itemize}

Recall that (1) $\mathcal{M}_{\phi}$ and $\mathcal{M}_{\psi}$ have identical architectures and are trained with different seeds. This pair represents independent-training method.

(2) $\mathcal{M}_{\phi_1}$, $\mathcal{M}_{\phi_2}$ are created from the same $\mathcal{M}_{\phi}$ by piecing together different rewriters $\mathcal{R}_{k_1}$ and $\mathcal{R}_{k_2}$ generated using different seeds. This pair represents parameter rewriting method.

(3) $\mathcal{M}_{\phi_1}$ and $\mathcal{M}_{\psi_2}$ are created by piecing $\mathcal{R}_{k_1}$ with $\mathcal{M}_{\phi}$ and piecing $\mathcal{R}_{k_2}$ with $\mathcal{M}_{\psi}$. This represents the scenario where parameter rewriting is applied to models which have already been independently trained.

\noindent
The local gradients are computed in the following way. For each image ${\textbf x}$, we compute:
\begin{itemize}[]
    \item $H_1 = \cos(\nabla_{{\textbf x}}  {\mathcal J}{\left(\phi, {\textbf x},{\textbf y}\right)},\nabla_{{\textbf x}}  {\mathcal J}{\left(\psi, {\textbf x},{\textbf y}\right)})$
    \item $H_2 = \cos(\nabla_{{\textbf x}}  {\mathcal J}{\left(\phi_1, {\textbf x},{\textbf y}\right)},\nabla_{{\textbf x}}  {\mathcal J}{\left(\phi_2, {\textbf x},{\textbf y}\right)})$
    \item $H_3 = \cos(\nabla_{{\textbf x}}  {\mathcal J}{\left(\phi_1, {\textbf x},{\textbf y}\right)},\nabla_{{\textbf x}}  {\mathcal J}{\left(\psi_2, {\textbf x},{\textbf y}\right)})$
\end{itemize}
where ${{\mathcal J}}(\cdot, \cdot,\cdot)$ is the training loss function and $\cos (\cdot, \cdot)$ is the cosine similarity function.
\begin{figure}[H]
  \centering
  \begin{subfigure}{\linewidth}
    \centering
    \includegraphics[width=0.7\linewidth]{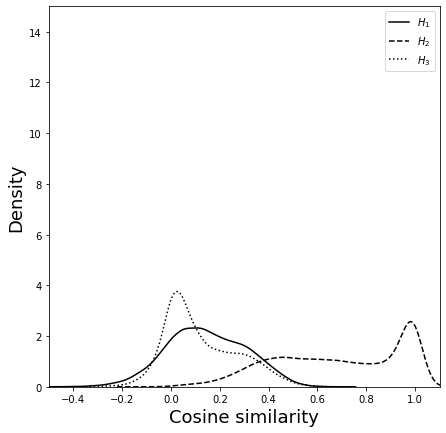}
    \caption{Attractor-based rewriter with spread spectrum decoder.}
    \label{fig:kde_spread}
  \end{subfigure}
  \begin{subfigure}{\linewidth}
    \centering
    \includegraphics[width=0.7\linewidth]{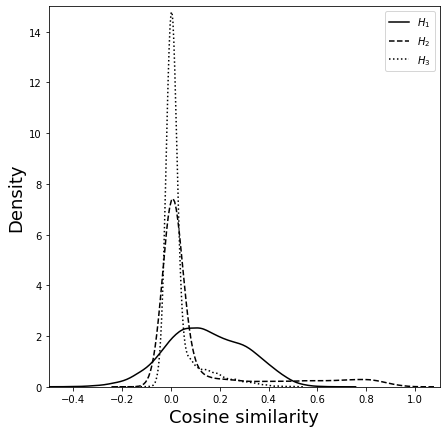}
    \caption{Attractor-based rewriter with QIM decoder.}
    \label{fig:kde_qim}
  \end{subfigure}%
  \vspace{-5pt}
    \caption{KDE plot of cosine similarity to show the effectiveness of applying attractor-based rewriters.}
    \label{fig:kde}
\end{figure}
\vspace{-15pt}
We conduct the same set of experiments using spread spectrum and QIM decoders respectively. The results are shown in Figure~\ref{fig:kde}. Based on the Kernel Density Estimate (KDE) plot, we found that two models trained with different seeds naturally have certain differences in local gradients (see $H_1$). Nonetheless, the mean cosine similarity is still quite high and the distribution is quite spread out, indicating that the adversary could potentially find a replicable adversarial sample relatively easily.

In Figure~\ref{fig:kde_spread}, attractor-based rewriter using spread spectrum decoder is applied. We can observe attractors changed the local properties such that the cosine similarity between local gradients spreads out and drops below $1$ (see $H_2$). However, a small portion of samples are left not covered by the attractors due to the limitation of spread spectrum and can be seen as the small peak near 1. When independent-training and parameter rewriting are combined (see $H_3$), the cosine similarity is brought closer to 0, indicating better performance. Overall, as spread spectrum only injects a very small number of attractors, the local gradient can only be changed to a limited extent.

In contrast, in Figure~\ref{fig:kde_qim}, when attractor-based rewriter using QIM decoder is applied, the local properties are significantly changed such that the cosine similarity between local gradients dropped drastically. This is observed as the peak of $H_2$ is sharp and is significantly higher than the peak of $H_1$ and the mean also shifts to near 0, showing very low similarity among the local gradients. $H_3$ has an even higher peak than $H_2$ near 0, indicating very strong performance when independent-training and rewriting methods are combined. Attractors injected by the QIM decoder scatter over the whole space and cover almost all samples, bringing the cosine similarity to almost 0. 

This experiment shows that different decoders can produce very different performance. An interesting future direction will be improving the proposed method using more effective watermarking schemes.

Note that most attacks have objective functions which either follow the gradients directly by obtaining backpropagated gradients or indirectly through some sampling or optimization techniques.  For example, consider an attack which uses the negative of training loss function $-{\mathcal J}(\cdot, \cdot,\cdot)$ as the attack objective function ${\mathcal L}(\cdot, \cdot,\cdot)$, so as to maximize the loss and cause misclassification. Suppose we can achieve:
 $$
\cos ( \nabla_{\textbf x} {\mathcal J} ( {\phi_1, {\textbf x}, {\textbf y}}) ,  \nabla_{\textbf x} {\mathcal J} ( {\phi_2, {\textbf x}, {\textbf y}})) \leq \mu
$$ then we can also achieve:
$$
\cos ( \nabla_{\textbf x} {\mathcal L} ( {\phi_1, {\textbf x}, {\textbf y}}) ,  \nabla_{\textbf x} {\mathcal L} ( {\phi_2, {\textbf x}, {\textbf y}})) \leq \mu
$$
Though the attack process starts from the same image, different local gradients in two copies lead the attack process to different adversarial regions in each copy, and this makes the generated adversarial samples non-replicable.

\subsection{Effectiveness of Binding (Permutation-based Rewriter)}
Unlike attractor-based rewriters which explicitly create new adversarial regions, permutation-based rewriter uses the binding mechanism to modify adversarial regions implicitly. In this section, we evaluate the effectiveness of the binding mechanism with respect to GeoDA~\cite{DBLP:conf/cvpr/RahmatiMFD20} attack (un-targeted) using the 10,000 testing images from CIFAR-10.

\begin{figure}[H]
    \centering
    \begin{subfigure}{\linewidth}
      \centering
      \includegraphics[width=0.7\linewidth]{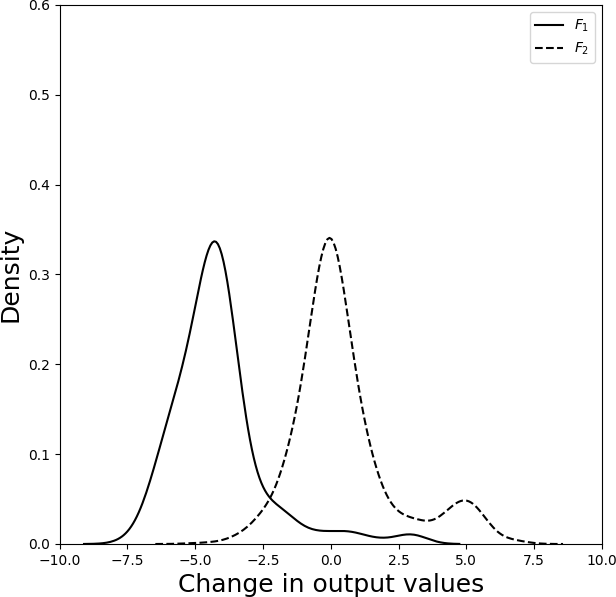}
      \caption{Original model without rewriter.}
      \label{fig:kde_ori}
    \end{subfigure}
    \begin{subfigure}{\linewidth}
      \centering
      \includegraphics[width=0.7\linewidth]{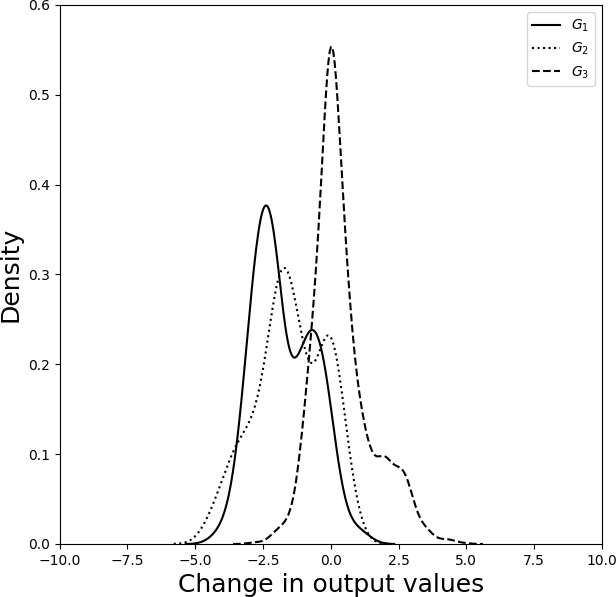}
      \caption{Model with permutation-based rewriter.}
      \label{fig:kde_per}
    \end{subfigure}
    \vspace{-5pt}
      \caption{KDE plot of change in output values caused by GeoDA attack to show the effectiveness of applying permutation-based rewriter.}
  \end{figure}

Firstly, we feed all samples into a randomly trained original model $\mathcal{M}_{\phi}$ and obtain all output values (before softmax). We then conduct GeoDA attack against $\mathcal{M}_{\phi}$ on each sample and record the new output values. We compute the differences between output values before and after the attack. In Figure~\ref{fig:kde_ori}, we plot two Kernel Density Estimate (KDE) curves: $F_1$ for change in output values for victim class $j$ (the original correct class or each image) and $F_2$ for all other classes. We then repeat the experiment on a model with permutation-based rewriter $\mathcal{M}_{\phi_1}$. In Figure~\ref{fig:kde_per}, we plot three KDE curves: $G_1$ for change in output values for victim class $j$, $G_2$ for binding partner of victim class ${\mathcal \pi}_{k_{1}}(j)$ and $G_3$ for all other classes.
\vspace{-8pt}
\vspace{-8pt}
In Figure~\ref{fig:kde_ori}, we can observe that $F_1$ has negative values while $F_2$ is almost symmetrically distributed around 0. This shows that when the un-targeted GeoDa moves away from the correct class (victim class), the prediction score of victim class decreases sharply while other classes remain mostly unchanged. In Figure~\ref{fig:kde_per}, both $G_1$ and $G_2$ have negative values with $G_2$'s mean slightly higher and peak slightly lower. $G_3$ has similar shape as $F_2$ in Figure~\ref{fig:kde_ori}. This shows that with the permutation-based rewriter, the change in prediction score is now split on victim class and its binding partner.

\subsection{Performance under Attack}
\label{sec:evaluation_under_attack}
We carried out boundary attack~\cite{DBLP:conf/iclr/BrendelRB18}, Hop Skip Jump attack (HSJ)~\cite{DBLP:conf/sp/ChenJW20} and Geometric Decision-based Attack (GeoDA)~\cite{DBLP:conf/cvpr/RahmatiMFD20}. We use Adversarial Robustness Toolbox (ART)~\cite{DBLP:journals/corr/abs-1807-01069} platform to conduct the experiments. 
We apply default setting of the toolbox.
For boundary attack, the maximum number of iteration is set to 5,000 with 25 trials per iteration.
For HSJ, the maximum number of iteration is 50 with 10,000 evaluations for estimating gradient. For GeoDA, the maximum number of iteration is set to 4,000.
\begin{table*}[t]
    \small
    \setlength{\tabcolsep}{1.5pt}
\centering
\begin{tabular}{|c|c|c|c|c|c|c|c|} 
    \hline
                                                                                                         & Datasets                        & \multicolumn{3}{c|}{CIFAR-10}                                                                  & \multicolumn{3}{c|}{GTSRB}                                                                      \\ 
    \hline
    Approaches                                                                                           & Attacks                         & \begin{tabular}[c]{@{}c@{}}Boundary\\Attack\end{tabular} & HSJ              & GeoDA            & \begin{tabular}[c]{@{}c@{}}Boundary\\Attack\end{tabular} & HSJ              & GeoDA             \\ 
    \hline
    \multirow{2}{*}{Independent Training}                                                                & Initial Attack Success Rate (a) & 96.8\% (2.0\%)                                         & 96.2\% (2.4\%) & 96.5\% (2.6\%) & 99.1\% (1.5\%)                                         & 99.0\% (1.6\%) & 99.2\% (1.8\%)   \\ 
    \cline{2-8}
                                                                                                         & Replication Rate (b)            & 5.5\% (0.6\%)                                          & 5.8\% (0.8\%)  & 27.0\% (1.9\%) & 5.4\% (0.6\%)                                          & 7.0\% (0.6\%)  & 7.8\% (0.9\%)   \\ 
    \hhline{|========|}
    \multirow{2}{*}{Permutation-based Rewriter}                                                          & Initial Attack Success Rate (c) & 95.8\% (2.2\%)                                         & 96.7\% (2.1\%) & 96.2\% (2.0\%) & 97.5\% (2.0\%)                                         & 97.5\% (1.7\%) & 97.8\% (1.9\%)  \\ 
    \cline{2-8}
                                                                                                         & Replication Rate (d)            & 22.7\% (0.9\%)                                         & 34.0\% (1.2\%) & 20.5\% (0.9\%) & 7.9\% (0.7\%)                                          & 19.8\% (0.9\%) & 12.1\% (1.0\%)  \\ 
    \hline
    \begin{tabular}[c]{@{}c@{}}Permutation-based Rewriter\\ + Training\end{tabular}                      & Replication Rate (e)            & 5.4\% (0.6\%)                                          & 4.5\% (0.6\%)  & 13.1\% (1.3\%) & 2.3\% (0.5\%)                                          & 2.7\% (0.6\%)  & 4.6\% (0.7\%)   \\ 
    \hhline{|========|}
    \multirow{2}{*}{\begin{tabular}[c]{@{}c@{}}Attractor-based Rewriter\\(Spread Spectrum)\end{tabular}} & Initial Attack Success Rate (f) & 96.3\% (1.9\%)                                         & 96.9\% (1.5\%) & 97.1\% (1.8\%) & 97.2\% (1.4\%)                                         & 98.1\% (1.5\%) & 98.4\% (1.4\%)  \\ 
    \cline{2-8}
                                                                                                         & Replication Rate (g)            & 40.1\% (1.0\%)                                         & 47.7\% (1.0\%) & 39.1\% (1.4\%) & 28.4\% (1.1\%)                                         & 27.5\% (0.6\%) & 13.5\% (0.7\%)  \\ 
    \hline
    \begin{tabular}[c]{@{}c@{}}Attractor-based Rewriter\\ + Training (Spread Spectrum)\end{tabular}      & Replication Rate (h)            & 5.2\% (0.6\%)                                          & 4.5\% (0.4\%)  & 21.5\% (0.8\%) & 2.7\% (0.6\%)                                          & 1.7\% (0.3\%)  & 3.3\% (0.2\%)   \\ 
    \hhline{|========|}
    \multirow{2}{*}{\begin{tabular}[c]{@{}c@{}}Attractor-based Rewriter\\(QIM)\end{tabular}}             & Initial Attack Success Rate (i) & 59.1\% (5.6\%)                                         & 97.5\% (1.5\%) & 94.8\% (1.4\%) & 79.5\% (1.4\%)                                         & 96.4\% (1.1\%) & 97.5\% (1.5\%)  \\ 
    \cline{2-8}
                                                                                                         & Replication Rate (j)            & 3.7\% (0.2\%)                                          & 2.5\% (0.2\%)  & 9.2\% (0.3\%)  & 0.7\% (0.2\%)                                          & 0.5\% (0.1\%)  & 8.4\% (0.4\%)   \\ 
    \hline
    \begin{tabular}[c]{@{}c@{}}Attractor-based Rewriter\\ + Training (QIM)\end{tabular}                  & Replication Rate (k)            & 3.7\% (0.2\%)                                          & 2.5\% (0.2\%)  & 5.3\% (0.3\%)  & 0.7\% (0.2\%)                                          & 0.1\% (0.1\%)  & 2.3\% (0.2\%)   \\
    \hline
    \end{tabular}
\caption{Performance of the proposed methods under attacks on CIFAR-10 and GTSRB dataset. The first value is the average attack success rate, and the second value in the bracket is the standard deviation.}
\label{tab:result}
\vspace{-15pt}
\end{table*}
The results are shown in Table~\ref{tab:result}. In our experiments, we look at the {\em initial attack success rate} and {\em replication rate on adversarial samples}.
They are described as follows:
\begin{itemize}[leftmargin=*]
    \item {\em Initial attack success rate.\ \ }
This measures the performance of adversarial attacks against a model. We used the testing dataset to conduct the attacks. We first feed all the samples into the model and gather only samples that are correctly classified. Then we apply the attack algorithms on these correctly classified samples. The attack is counted as successful if the generated output gets misclassified into any class other than the correct class (i.e. un-targeted attack). 
\item {\em Replication rate on adversarial samples.\ \ }
This is percentage of adversarial samples on one model $\mathcal{M}$ that are successfully replicated on another mode $\mathcal{M'}$. That is, suppose ${\textbf A}$ is the set of  samples generated by the attack on $\mathcal{M}$, ${\textbf B}\subseteq {\textbf A}$ is the set of adversarial samples that is misclassified by $\mathcal{M}$ and ${\textbf C}\subseteq {\textbf A}$ is the set of adversarial samples that is misclassified by $\mathcal{M'}$, then the rate is $|{\textbf B}\cap{\textbf C}|/|{\textbf B}|$.
\end{itemize}
\noindent
{\bf Independent Training.}\ \
For each dataset, 10 different models are trained independently with the same architecture from scratch using different random seeds.  We treat one model as the adversary’s copy $\mathcal{M}_{\phi}$ and another as the victim’s copy $\mathcal{M}_{\psi}$. Hence, we have $P_{2}^{10}=90$ pairs of ($\mathcal{M}_{\phi}$,$\mathcal{M}_{\psi}$) for evaluation. We report the average initial attack success rate and average replication success rate in rows (a) and (b) respectively. The average and standard deviation are computed according to number of models/pairs, for instance, among 10 copies in (a) and among 90 pairs in (b).
\\ \\[-3pt]
\noindent
{\bf Parameter Rewriting.}\ \
For each parameter rewriting method, we start from one independently trained base model $\mathcal{M}_{\phi}$ and create 10 copies using random seeds. We can obtain $P_{2}^{10}=90$ pairs (of $\mathcal{M}_{\phi_1}$ and $\mathcal{M}_{\phi_2}$) for evaluation. As we mentioned above, we have 10 independently trained base models, so we repeat the evaluation for each base model.
The average results are reported in row (c) and (d) for permutation-based rewriter, rows (f) and (g) for attractor-based rewriter using spread spectrum decoder, and rows (i) and (j) for attractor-based rewriter using QIM decoder. The average and standard deviation are computed for $10 \times 10=100$ copies in row (c), (f) and (i) and for $90 \times 10=900$ pairs in row (d), (g) and (j).
\\ \\[-3pt]
\noindent
{\bf Training + Rewriting.}\ \
We reuse the 10 independently trained base models to create $P_{2}^{10}=90$ attacker-victim pairs and apply parameter rewriting to rewrite each ($\mathcal{M}_{\phi}$,$\mathcal{M}_{\psi}$) pair to ($\mathcal{M}_{\phi_1}$,$\mathcal{M}_{\psi_2}$) using different random seeds then conduct the evaluation.
The average results are reported in row (e) for permutation-based rewriter, rows (h) for attractor-based rewriter using spread spectrum decoder, and rows (k) for attractor-based rewriter using QIM decoder.

From the results, we observe independent-training is able to lower success rate for boundary attack and HSJ. However, its performance against GeoDA is not optimal in CIFAR10. 
On the other hand, parameter rewriting methods also can significantly lower the success rate of replication attack.
Among the three constructions we created for parameter rewriting, attractor-based rewriter using QIM decoder has the best performance and can achieve even lower attack success rate than expensive independent-training. When parameter rewriting is applied to models obtained through independent-training (combing independent-training and parameter rewriting methods), both permutation-based approach and attractor-based approach can consistently lower the success rate of replication attack.
\vspace{-10pt}

\begin{table}[H]
    \centering
    \small
    \setlength{\tabcolsep}{3pt}
    \begin{tabular}{|c|c|c|c|} 
    \hline
                                                                  & \begin{tabular}[c]{@{}c@{}}Independent\\Training\end{tabular}                                   & \begin{tabular}[c]{@{}c@{}}Attractor-based\\Rewriter (QIM)\end{tabular} & \begin{tabular}[c]{@{}c@{}}Permutation-based\\Rewriter\end{tabular}  \\ 
    \hline
    \begin{tabular}[c]{@{}c@{}}Additional\\Training\end{tabular}  & \begin{tabular}[c]{@{}c@{}}3,800 s\\per copy~\footnotemark[1]\end{tabular}                                       & Not required                                                                & Not required                                                                   \\ 
    \hline
    \begin{tabular}[c]{@{}c@{}}Additional\\Storage~\footnotemark[2]\end{tabular}   & \begin{tabular}[c]{@{}c@{}}78,366 KB\\per copy\end{tabular} & \begin{tabular}[c]{@{}c@{}}2 KB\\per copy\end{tabular}            & Not required                                                                   \\ 
    \hline
    \begin{tabular}[c]{@{}c@{}}Additional\\Inference\end{tabular} & Not required                                                                                              & 0.4 ms                                                            & Not required                                                                   \\
    \hline
    \end{tabular}
    \caption{Additional cost of independent-training and parameter rewriting.}
    \label{tab:cost}
    \vspace{-15pt}
\end{table}
\footnotetext[1]{\color{black} Independent-training method trains each copy from scratch to ensure randomness.}
\footnotetext[2]{\color{black} For independent-training, additional storage refers to storage space (model size) incurred by each additional distributed copy. For attractor-based rewriter, it refers to the size of watermark in each distributed copy.} 
Note the cost of above approaches are different. For independent-training, each copy needs additional resource and time for training, and additional storage to store the model weights. Attractor-based approach needs additional storage to store the seeds of rewriter and the rewriter component also adds a bit of overhead during inference. Permutation-based rewriter has virtually no cost. In Table~\ref{tab:cost}, we show the additional cost of each approach when we conduct the evaluation using VGG-19 on CIFAR-10. The training and inference are conducted on an NVIDIA Tesla V100. Training the original copy takes around 3,800 seconds (200 epochs) on average. The inference takes 3.5 ms on average (batch size 1).

\subsection{Collusion Attack}
\label{sec:collusion}
The adversary may purchase or subscribe to the model under multiple names and gain black-box access to multiple copies. With accesses to multiple copies, the adversary potentially could launch a {\em collusion attack} that collates information from different copies.

This attack is similar to collusion in digital watermarking, where the attacker attempt to remove watermarks by combining multiple watermarked copies of the same host~\cite{DBLP:journals/tip/WangWZTL05}. Intuitively, access to multiple black-boxes could allow the adversary to compare and aggregate different models' outputs on the same input, thus reduce the influence of parameter rewriting.

Since our framework is new, there is currently no known collusion attack. Therefore, for preliminary investigation, we designed a collusion attack based on DeepFool~\cite{MoosaviDezfooli2016DeepFoolAS}. In our setup, the adversary uses the set of available copies as an ensemble. The ensemble aggregates the output of each individual copy and returns the mean as the final output. The DeepFool attack is mounted against the ensemble model to generate adversarial samples. As DeepFool generates perturbation by finding the nearest hyperplane of an input, it now searches the nearest hyperplane in the ensemble model, instead of an individual copy. In such a way, individual copys' decision boundary cancel each other out inside the ensemble. The differences of copies induced by parameter rewriting are suppressed.

In addition, for the adversarial samples generated by the above DeepFool attack, the adversary only keeps samples that are able to attack every individual copy in the ensemble and use them for the replication attack against the victim copy. In other words, when we evaluate the ``success rate by applying adversarial samples'', we require the adversarial sample to not only fool the ensemble model but also every individual copy in the ensemble.

Note that DeepFool is a white-box attack. As there are methods that ``reverse-engineer'' the parameters using techniques such as model cloning~\cite{DBLP:conf/uss/TramerZJRR16,DBLP:conf/ccs/PapernotMGJCS17,DBLP:series/lncs/OhSF19,DBLP:conf/sp/WangG18,DBLP:conf/kdd/LowdM05,DBLP:conf/fat/MilliSDH19,DBLP:conf/cvpr/OrekondySF19}, we assume the adversary is powerful and can retrieve all parameters through reverse-engineering. Hence, the white-box attack here can serve as a bound on black-box attacks that incorporate an effective cloning where the attacker successfully converts the victim from black-box to white-box.

\vspace{-10pt}
\begin{figure}[H]
    \begin{subfigure}{\linewidth}
      \centering
      \includegraphics[width=0.73\linewidth]{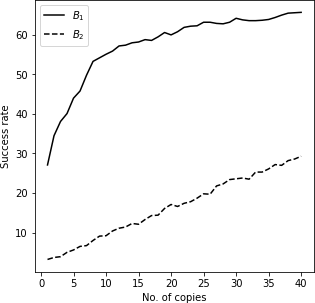}
      \caption{Success rate by applying all generated samples.}
    \end{subfigure}
    \begin{subfigure}{\linewidth}
      \centering
      \includegraphics[width=0.73\linewidth]{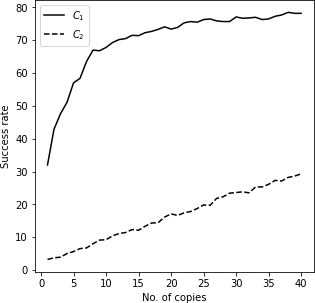}
      \caption{Success rate by applying only adversarial samples.}
    \end{subfigure}
    \vspace{-10pt}
    \caption{Performance of collusion attack (Attractor-based).}
    \label{fig:collusion}
\end{figure}
\vspace{-5pt}

We test collusion attack in two settings:
\begin{enumerate}[leftmargin=*]
    \item All the models are distributed using the independent-training method. The malicious buyer has black-box access to a set of copies with the same architecture but trained with different seeds. The outputs generated by the attack is then applied to the victim's copy.
    \item In addition to the independent-training step in setting (1), we apply attractor-based rewriter with QIM decoder. 
\end{enumerate}

We denote copies obtained through the independent-training method as $\mathcal{M}_{\zeta_1}, \allowbreak \mathcal{M}_{\zeta_2},\cdots,\mathcal{M}_{\zeta_r}$, and we denote these copies obtained through independent-training combined with attractor-based parameter rewriting as $\mathcal{M}_{\eta_1},\mathcal{M}_{\eta_2},\cdots,\mathcal{M}_{\eta_r}$. In every subfigure, the solid line represents the rate under setting 1, and the dotted line represents the rate under setting 2. We plot following values:

\begin{itemize}[leftmargin=*,topsep=0pt]
    \item $B_1$: Replication rate by applying all outputs generated from\\ $\{\mathcal{M}_{\zeta_1},\mathcal{M}_{\zeta_2},\cdots,\mathcal{M}_{\zeta_r}\}$ on victim's copy $\mathcal{M}_{\zeta_{r+1}}$
    \item $B_2$: Replication rate by applying all outputs generated from\\ $\{\mathcal{M}_{\eta_1},\mathcal{M}_{\eta_2},\cdots,\mathcal{M}_{\eta_r}\}$ on victim's copy $\mathcal{M}_{\eta_{r+1}}$
    \item $C_1$: Replication rate by applying adversarial samples generated from $\{\mathcal{M}_{\zeta_1},\mathcal{M}_{\zeta_2},\cdots,\mathcal{M}_{\zeta_r}\}$ on victim's copy $\mathcal{M}_{\zeta_{r+1}}$
    \item $C_2$: Replication rate by applying adversarial samples generated from $\{\mathcal{M}_{\eta_1},\mathcal{M}_{\eta_2},\cdots,\mathcal{M}_{\eta_r}\}$ on victim's copy $\mathcal{M}_{\eta_{r+1}}$
\end{itemize}

In Figure~\ref{fig:collusion}, we observe that the success rate of replication attack indeed increases with the increasing number of available copies. The influence of attractors injected using QIM decoder decreases linearly. Nonetheless, it is still a significant improvement over distributing identical copies or directly applying the independent-training. Using an anti-collusion watermark~\cite{DBLP:journals/tit/SkoricVCT08} scheme to prevent collusion attacks will be interesting future work.

\vspace{8pt}
\section{Conclusion and Future Work}
Adversarial attacks are difficult to defend.
In this paper, we sidestep the general setting and consider a weaker form of attacks where the attacker's goal is to replicate adversarial attack from one copy to another.  We proposed an active approach to rewrite parameters of different copies as a measure to prevent replication. We constructed two different types of rewriters: attractor-based rewriter and permutation-based rewriter. Our evaluation demonstrates promising results and we believe that there are a number of interesting directions to explore.  For instance, incorporation of anti-collusion fingerprinting code to address collusion attack or using more suitable watermarking decoder in attractor-based rewriter for better performance. 

One important future work will be applying the proposed method to other non-image data types such as audio and text. 
The proposed permutation-based rewriting method manipulates the output of a classification model regardless of the input types. Thus, it can be easily applied to other data types. The attractor-based method relies on a watermark decoder. Hence, for non-image data type, one could adopt existing watermarking scheme on that data type (if any) for parameters rewriting.

In addition, the proposed approach could also be used for attack tracing and attack deterrence.

\begin{itemize}[leftmargin=*,topsep=0pt]
\item {\em Attack Tracing.\ \ }
    By controlling the model distributive process, the seller could also potentially trace the origin of adversarial samples. For example, in the attractor-based rewriter, different attractors are injected into different copies of the model, and each copy will have different local properties. Thus, it is analogous to have a distinct model signature of each copy. Any adversarial sample generated from a particular copy would inadvertently carry a unique imprint. Suppose the seller keeps a record of the signatures of the multiple distributed copies, it is possible to analyze an adversarial sample and ultimately trace it back to the model that produced that sample. 

\item {\em Attack Deterrence.\ \ } 
    Seller's ability to trace the attack origin can also deter potential attackers. The adversary will probably need to design a more sophisticated attack which can achieve the attack target while protecting attacker's identity.
\end{itemize}
\clearpage

\bibliographystyle{ACM-Reference-Format}
\bibliography{paper}

\end{document}